\documentclass[fleqn,usenatbib]{mnras}
\usepackage{newtxtext,newtxmath}
\usepackage[T1]{fontenc}
\usepackage{ae,aecompl}
\usepackage{dirtytalk}

\usepackage{graphicx}	% Including figure files
\usepackage{amsmath}	% Advanced maths commands
\usepackage{amssymb}	% Extra maths symbols
\usepackage[dvipsnames]{xcolor}
\hypersetup{colorlinks, citecolor = Aquamarine}
\title[Observed glitches in 8 young pulsars]{Observed glitches in 8 young pulsars}

\author[A. Basu et al.]{
Avishek Basu$^{1}$\thanks{E-mail: avishek@ncra.tifr.res.in}\href{https://orcid.org/0000-0002-4142-7831}{\includegraphics[scale=.5]{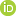}},
Bhal Chandra Joshi$^{1}$\href{https://orcid.org/0000-0002-0863-7781}{\includegraphics[scale=.5]{ORCID-iD_icon-16x16.png}},
M. A. Krishnakumar$^{2}$\href{https://orcid.org/0000-0003-4528-2745}{\includegraphics[scale=.5]{ORCID-iD_icon-16x16.png}},
\newauthor 
Dipankar Bhattacharya$^{3}$,
Rana Nandi$^{4}$\href{https://orcid.org/0000-0002-6277-2618}{\includegraphics[scale=.5]{ORCID-iD_icon-16x16.png}},
Debades Bandyopadhyay$^{5}$\href{https://orcid.org/0000-0003-0616-4367}{\includegraphics[scale=.5]{ORCID-iD_icon-16x16.png}},
\newauthor 
Prasanta Char$^{6}$\href{https://orcid.org/0000-0001-6592-6590}{\includegraphics[scale=.5]{ORCID-iD_icon-16x16.png}},
and P. K. Manoharan$^{1,7}$\href{https://orcid.org/0000-0003-4274-211X}{\includegraphics[scale=.5]{ORCID-iD_icon-16x16.png}}
\\
\\
$^{1}$National Centre for Radio Astrophysics-Tata Institute of Fundamental Research, Ganeshkhind, Post-Bag 3 , Pune 411007, India\\
$^{2}$Fakult\"{a}t f\"{u}r Physik, Universit\"{a}t Bielefeld, Postfach 100131, 33501 Bielefeld, Germany\\
$^{3}$Inter University Centre for Astronomy and Astrophysics, Post Bag 4, Pune 411007, India\\
$^{4}$ Department of Nuclear and Atomic Physics, Tata Institute of Fundamental Research, Mumbai 400005, India\\
$^{5}$ Astroparticle Physics and Cosmology Division, Saha Institute of Nuclear Physics, HBNI, 1/AF Bidhannagar, Kolkata-700064, India\\
$^{6}$ INFN Sezione di Ferrara, Via Saragat 1, I-44122 Ferrara, Italy\\
$^{7}$ Radio Astronomy Centre, NCRA-TIFR, Udagamandalam, India
}

\date{Accepted 2019 November 12. Received 2019 November 11; in original form 2019 September 2}

\pubyear{2015}

\begin{document}
\label{firstpage}
\pagerange{\pageref{firstpage}--\pageref{lastpage}}
\maketitle

\begin{abstract}
The abrupt change in the pulse period of a pulsar is called a pulsar glitch. 
%So far 549 such glitches in 189 pulsars have been reported in the Jodrell Bank Glitch catalogue. 
In this paper, we present eleven pulsar glitches detected using the Ooty Radio Telescope (ORT) and the upgraded Giant Metrewave Radio Telescope (uGMRT) in high cadence timing observations of 8 pulsars.  
%The glitches reported here were detected in 8 of these 18 pulsars during 2017-2019. 
The measured relative amplitude of glitches ($\Delta \nu/\nu$) from our data ranges from $10^{-6}$ to $10^{-9}$. Among these glitches, three are new discoveries, being reported for the first time. We also reanalyze the largest  pulsar glitch in the Crab pulsar (PSR J0534+2200) by fitting the ORT data to a new phenomenological model including  the slow rise in the post-glitch evolution. We measure an exponential recovery of 30 days after the Vela glitch detected on MJD 57734 with a healing factor $Q=5.8\times 10^{-3}$. Further, we report the largest glitch ($\Delta \nu/\nu = 3147.9 \times 10^{-9}$) so far in PSR J1731$-$4744.
\end{abstract}


%%%%%%%%%%%%%%%%%%%%%%%%%%%%%%%%%%%%%%%%%%%%%%%%%%

%%%%%%%%%%%%%%%%% BODY OF PAPER %%%%%%%%%%%%%%%%%%

\section{Introduction}

Pulsars are rapidly rotating magnetized neutron stars (NS). Their rotation frequency ($\nu$) decreases with time due to loss of rotational kinetic energy by radiation and particle wind. Other than the slow systematic spin down, pulsars occasionally spin up. The sudden spinning up of a pulsar is called a glitch \citep{Radman69}. The relative spin up $\frac{\Delta \nu}{\nu}$ spans a few orders of magnitude from $10^{-10}$ to $10^{-6}$. This sudden spin up of the pulsar can be measured by the technique of pulsar timing \citep{sl96}. 
%\citep{taylor1992,sl96}. 
%The region of the pulsed radio emission corotates with the star. Hence, any disturbance in the star's rotational motion can be tracked by measuring the radio pulse time of arrival (ToA). While glitch is one kind of timing irregularity, pulsars also exhibit another type of deviation from stable rotation named as timing noise \citep{cor80}. Timing noise is a slow systematic wandering of time of arrival (ToA) residuals. 
Initially, glitches were modelled as starquakes \citep{ruderman69}. But the frequent glitches in the Vela pulsar favoured a superfluid model \citep{AlparvortexcreepI1984,AlparvortexcreepII1984,alparpines1985,BaymEipstein1988,haskellmelatos15} over the starquake model. In any case, glitches provide a unique probe of the structure of NS. 

In the superfluid origin of glitches, they are caused by the sudden release of accumulated angular momentum of the vortices pinned in the inner crust \citep{AndersonItoh1975} when the difference in the angular velocity between the superfluid (SF) and its surroundings exceeds a critical value.  Migration and re-pinning of vortices over an extended period of time defines the post-glitch behaviour. Thus, glitches can be used to probe the structural properties of neutron stars \citep{link1992} and also the dynamics of SFs inside the neutron star \citep{Seveso2012snowplough}. 
The moment of inertia (I) of the star and that of the crustal SF is determined by its equation of state (EoS) and hence fractional moment of inertia (FMI) can be used to constrain the EoS. However, in some cases the observed FMI is so large that the participation of the core SF appears to be necessary \citep{basuFMI}. The contributions of different components of NS to glitches is still an open problem. 

In this paper, we present the results of recent glitches that have been detected using the Ooty Radio Telescope \citep[ORT: ][]{swarup1971large} and the upgraded Giant Metrewave Radio Telescope \citep[uGMRT: ][]{sak+91,gak+17} in a program of high cadence monitoring of high glitch rate pulsars. The ORT is a 530 m long offset parabolic cylindrical antenna in the north-south direction and 30 m wide in the east-west direction. The ORT is sensitive to a single polarization of the incoming radiation.
The uGMRT is a \say{Y} shaped interferometer with 30 antenna elements, each of 45 m in diameter.
\\

The main aim of our targeted observations was to detect large number of glitches and find the post-glitch recovery with high cadence monitoring of a selected sample of pulsars. The organisation of the paper is as follows. In Section \ref{smapleselection}, we discuss the method by which we have selected the pulsars to create our sample set for targeted observations. In Section \ref{observations+analysis}, we discuss the observation set up used for observing the pulsars at the uGMRT and the ORT as well as the analysis procedure to reduce the raw data. Finally, in Section \ref{results}, we present the glitch parameters of individual pulsars from our timing studies and conclude the paper with discussion in Section \ref{end}.

\section{Sample selection and Integration Time}
\label{smapleselection}

We selected a sample of pulsars, visible at the ORT and the uGMRT, with the following considerations in mind. First, we used the catalogue of pulsar glitches, maintained by Jodrell Bank\footnote{http://www.jb.man.ac.uk/pulsar/glitches.html} to select the glitching pulsars, which were visible at the ORT and the uGMRT. Thus, all pulsars with declinations from $-55^{\circ}$ to $+55^{\circ}$ were selected for observations with the ORT. Likewise all sources with declinations from $-53^{\circ}$ to $+90^{\circ}$ were selected for observations at the uGMRT. Second, the telescope and the observing frequency of each pulsar was decided based on its Dispersion Measure (DM)\footnote{Dispersion measure is integrated electron density towards the line of sight of the pulsar, which causes the higher frequency pulse to arrive before the lower frequency pulse.}. As the ORT operates at 334.5 MHz, the pulsars with DMs greater than 124 pc cm$^{-3}$ (with the exception of PSR J1740$-$3015) were not observed with the ORT. This is because the extreme scatter broadening at lower frequency of operation at the ORT results in degradation of the signal to noise ratio (S/N). In contrast, all the high DM pulsars were selected to be observed at either 750 or 1300 MHz using the uGMRT. 

Then, the list was screened for detectabilty of pulsars with the ORT and/or the uGMRT as follows. For timing these pulsars, a minimum S/N of 20 was assumed to be adequate as an optimal trade-off between the required telescope observing time and timing precision for these timing noise dominated pulsars. The required observations time $\tau$ was then estimated using the radiometer equation, given below

\begin{eqnarray}
\label{Modifiedradpulsar}
\tau = \frac{(S/N)^2 (T_{rec}+T_{sky})^2}{G^2 S_{average}^2N_p \Delta f}\Big(\frac{W}{P-W}\Big)
\end{eqnarray}

where $T_{rec} \, \text{and}\,  T_{sky}$ are receiver temperature and sky temperature. The $T_{sky}$ is computed from the Haslam map \citep{haslammap1982} assuming the  spectral index for sky background to be  2.6 as mentioned in \citet{haslammap1982}. $G$, $\Delta f$ and $N_p$  are the gain of the telescope, the bandwidth of the observations and the number of polarization respectively. While the ORT has a single polarization, the uGMRT has dual  polarization feeds. $P$ (pulse period) was taken from the ATNF pulsar catalog\footnote{http://www.atnf.csiro.au/research/pulsar/psrcat/}, whereas $W$ (width of the pulse) was corrected for pulse scatter-broadening  from $W_{50}$\footnote{$W_{50}$ is the width of the pulse at the 50\% of the peak, obtained from ATNF pulsar catalogue} as explained below. The effect of scatter broadening was taken into account by convolving a top hat pulse of width $W_{50}$ with the exponential scatter-broadening function using characteristic time-scales obtained from \citet{kk17} and NE2001 model \citep{NE2001}. 
%The normalization of the convolved pulse profile are done by conserving the area under the curve before and after the convolution. 
\begin{table*}
\begin{tabular}{|l|l|l|l|l|l|l|l|}
\hline
\hline
PSR J-Name & 
\begin{tabular}[c]{@{}l@{}} RA\\(h:m:s)\end{tabular} & 
\begin{tabular}[c]{@{}l@{}} Dec. \\($\circ:\prime:\prime \prime$)\end{tabular} &
\begin{tabular}[c]{@{}l@{}}Start\\ MJD\end{tabular} & 
\begin{tabular}[c]{@{}l@{}}Stop\\ MJD\end{tabular} & 
\begin{tabular}[c]{@{}l@{}}Cadence\\ (per week)\end{tabular} & 
\begin{tabular}[c]{@{}l@{}}Duration\\(min)\end{tabular} & 
\begin{tabular}[c]{@{}l@{}}Ref \\position\end{tabular} \\ \hline \hline
             &  & &       &  \textbf{ORT} &   &       &         \\ \hline
J0534$+$2200 &05:34:31.973&+22:00:52.06& 56729 & 58323 & ~~~~~~~7 & ~~~~~~~5 & ~~~~~~~5   \\ \hline
J0742$-$2822 &07:42:49.058&-28:22:43.76& 56630 & 57870 & ~~~~~~~7 & ~~~~~~~10 & ~~~~~~~7  \\ \hline
J0835$-$4510 &08:35:20.61149&-45:10:34.8751& 56729 & 58299 & ~~~~~~~7 & ~~~~~~~5 & ~~~~~~~6    \\ \hline 
J1731$-$4744 & 17:31:42.17 & -47:44:34.37 & 57919 & 58299& ~~~~~~~2 & ~~~~~~~20 & ~~~~~~~1\\ \hline
J1740$-$3015 &17:40:33.82 &-30:15:43.5 & 56779 & 58312 & ~~~~~~~1 & ~~~~~~~180& ~~~~~~~1\\ \hline
%J1825$-$0935 &  & & 57820 & 58321 & 2 & & 15  \\ \hline \hline
             &  & &       &  \textbf{uGMRT} &   &       &       \\ \hline \hline
J0729$-$1448(4) &07:29:16.45&-14:48:36.8& 58097 & 58663 & ~~~0.5 - 0.2 & ~~~~~~~20 & ~~~~~~~1 \\ \hline
J1740$-$3015(5) &17:40:33.82 & -30:15:43.5& 58089 & 58455 & ~~~0.5 - 0.2 & ~~~~~~~15 &~~~~~~~1  \\ \hline 
J1751$-$3323(4) & 17:51:32.725 & -33:23:39.6& 58097 & 58663 & ~~~0.5 - 0.2 & ~~~~~~~20 &~~~~~~~4\\ \hline 
J1837$-$0604(5) &18:37:43.55&-06:04:49 & 58089 & 58455 & ~~~0.5 - 0.2& ~~~~~~~5 & ~~~~~~~3  \\ \hline \hline
%J1952$+$3252(5) &19:52:58.206&+32:52:40.51& 58089 & 58671 & 0.5 - 0.2& 2 & 15 \\ \hline \hline
\end{tabular}
\caption{Table of pulsars, which exhibited a glitch, in our monitoring program with the ORT and the uGMRT. Observations at the ORT were done at 334.5 MHz with 16 MHz bandwidth. At uGMRT, observations were done at Band 5 (1100$-$1500 MHz) and Band 4 (550$-$950 MHz), with 400 MHz bandwidth. The first column shows the J Name of the pulsar, second and third column contains the RA and Dec taken from  references listed in column 8. In case of pulsars observed at uGMRT, the number in the bracket adjacent to the name of the pulsar shows the band used during the observations. The fourth and fifth column show the start and stop MJD of the observations between which the glitches have been detected. The second last column tabulates the on-source integration time. 
\newline References :1. \citet{Yu2013}, 2. \citet{zbcg08}, 3. \citet{DAmico2001}, 4. \citet{Yuan2010}, 5. \citet{McNamara71}, 6. \citet{dlrm03}, 7. \citet{HLK2004}}
\label{Obs_tab}
\end{table*}
The final width $W$ is taken as 50\% of the peak height of the convolved pulse profile. $S_{average}$ is the average flux density of the pulsar, which is specified in the ATNF catalog at either 400 and/or 1400 MHz. We computed the $S_{average}$ of the pulsar at the observing frequency $f$ using the relation $F_{f} \propto f^{-\alpha}$, where $\alpha$ is the spectral index. For the pulsars, whose flux densities are known at two frequencies in the  ATNF pulsar catalogue, we compute the spectral index ($\alpha$) from the values quoted in the catalogue.  Otherwise,  we used $\alpha = 1.8$ \citep{mkkw00} to scale $S_{average}$ from the flux density at the frequency available in the catalog. Then, we used Equation \ref{Modifiedradpulsar} to estimate the required integration time for each pulsar at the ORT and at the uGMRT.  The pulsars whose integration time was more than 40 minutes were not selected. One exception was higher cadence observations of PSR J1740$-$3015 using the ORT as this pulsar is heavily scatter-broadened at 334 MHz. Observation time required at the uGMRT was about 15 minutes at 1300 MHz. The final integration time was determined (even for the pulsars, which required very short integration times)  by the time required to accumulate 2000 pulses. This enabled us to obtain at least two sub-integrations each of 1000 pulses in every epoch of observations for all the pulsars. These considerations reduced our sample of pulsars to 29. Finally, the glitch rate per year was 
used to narrow down the list further. The pulsars having a glitch rate of more than 0.3 per year were selected for observations. In  pilot observations, we detected only 18 of them in the estimated integration time and these were monitored for the study presented here. Only 8 out of these 18 pulsars were detected with glitches. This sample covers a wide range of characteristic ages between 1.26 kyr to 984 kyr. The cadence and observations are discussed in the Section \ref{observations+analysis}.

\section{Observations and Analysis}
\label{observations+analysis}
Observations of the selected sample of pulsars were carried out with  the ORT and the uGMRT as summarized in Table \ref{Obs_tab}. The observations at the ORT were carried out at 334.5 MHz with 16 MHz bandwidth. The pulsar back end PONDER \citep{njmk15}  was used during the observations at the ORT. PONDER provides  real-time coherently dedispersed time series data of the pulsar.  Folding of the dedispersed time series data was done using an ephemeris created from our initial observations using pulsar analysis software DSPSR\footnote{http://dspsr.sourceforge.net/} \citep{vsm11} and this provided  folded sub-integrations of about 1000 pulses as PSRFITS \citep{hvsm04} files. The time stamps for the data were  provided in Coordinated Universal Time (UTC) obtained using a Global Positioning System receiver. 

For the pulsar observations at uGMRT, we used the 14 central compact array antennas and the first antenna of each arm in a phased array mode. Phasing of the array is achieved by observing a standard calibrator source in the sky before the pulsar observations and compensating for instrumental delays in the receiver chain of each antenna estimated from these observations. Observations of the pulsars were done at Band 4 (550 $-$ 950  MHz)  and Band 5 (1100 $-$ 1500 MHz) with 400 MHz bandwidth at the uGMRT with two channels of polarisation. The data were recorded with  2048 channels and processed offline. The native uGMRT format spectral data  were converted to SIGPROC\footnote{http://sigproc.sourceforge.net/} format filterbank file, which was dedispersed and folded using DSPSR to obtain PSRFITS files with sub-bands and sub-integrations. Ephemerides, obtained from initial solutions of our data, were used for folding the time-series. The time stamps for the data were  provided in UTC obtained using a Global Positioning System receiver. 

\begin{table*}
\begin{tabular}{llllllll}
\hline \hline
\multicolumn{1}{|l|}{\begin{tabular}[c]{@{}l@{}}Pulsar \\J Name\end{tabular}} & \multicolumn{1}{l|}{\begin{tabular}[c]{@{}l@{}}Epoch \\ MJD\end{tabular}} & \multicolumn{1}{l|}{\begin{tabular}[c]{@{}l@{}}Preglitch \\ $\nu$ (Hz)\end{tabular}}&
\multicolumn{1}{l|}{\begin{tabular}[c]{@{}l@{}}Preglitch \\ $\dot{\nu}$ (Hz s$^{-1}$)\end{tabular}}&
% \multicolumn{1}{l|}{\begin{tabular}[c]{@{}l@{}}Postglitch \\ $\nu$ (Hz)\end{tabular}}&
%\multicolumn{1}{l|}{\begin{tabular}[c]{@{}l@{}}Postglitch \\ $\dot{\nu}$ (Hz/s)\end{tabular}}&
\multicolumn{1}{l|}{$\frac{\Delta \nu}{\nu} \times 10^{-9}$} & 
\multicolumn{1}{l|}{$\frac{\Delta \Dot{\nu}}{\Dot{\nu}}  \times 10^{-3}$} & 
\multicolumn{1}{l|}{$\tau$ (days)} &
\multicolumn{1}{l|}{$\frac{\Delta \text{I}}{\text{I}}(\%)$}
 \\ 
\hline \hline
J0534$+$2200 & 58064.9(1) & 29.636731054(1) & -3.68607(4) E-10 & 484.39(1) & 5.173(6)&$--$&0.2\\ \hline
J0534$+$2200 &  58237.2(1) & 29.631219537(6) & -3.6899(2) E-10 & 1.7(6)& -0.144(8)&$--$&0.0009\\ \hline
J0729$-$1448 & 58266.6(9) & 3.9725934652(4)& -1.79807(2) E-12 & 3.8(4) & $--$ & $--$&0.003 \\ \hline
J0742$-$2822 &  56726.1(2)  & 5.9961735068(4) & -6.042(2) E-13 & 2.6(2) & $--$ &$--$&0.017 \\ \hline
J0835$-$4510 &  57734.4(2) & 11.186433252(6)  & -1.556383(8)E-11 &1433.2(9)& 5.595(9)&32(2)&1.45\\ \hline
J1731$-$4744 & 57978.17(2)& 1.204923648(2)& -2.373(1) E-13 & 3147.9(2)& 2.2(4) &$--$&4.2\\ \hline 
%J1740$-$3015 & 58435(2)& 1.6473610601(3)& -1.26599(2) E-12& 3(2) & $--$ & $--$ \\ \hline
J1740$-$3015 & 57336(9) &  1.64747945(1) & -1.2637(4)E-12&4(2)& -0.6(8)& $--$&0.004 \\ \hline
J1740$-$3015 & 57468.8(5) & 1.64745987(1) & -1.263(2)E-12 & 235(28)& $--$& $--$&2.66 \\ \hline
J1740$-$3015 & 58236.0(3) & 1.6473814477(1) & -1.26412(2)E-12 & 837.4(2) & 1.5(6)   & $--$&1.64 \\ \hline
J1751$-$3323 & 58438(3) & 1.8240514978(2) & -3.019(2) E-14& 3.4(8) & 27(4)&$--$&0.17 \\ \hline
J1837$-$0604 & 58233.7(8) & 10.382114964(4)& -5.025(2) E-12 & 70(1) & $--$ & $--$&0.1 \\ \hline
\end{tabular}
\caption{This table lists the glitch parameters for all the glitches presented in this work. The first column is the J name of the pulsars, followed by the epoch of the glitch, pre-glitch rotation frequency and the frequency derivative at the glitch epoch. In the fifth, sixth and the seventh columns, we present  the fractional change in rotation frequency and its derivative due to glitch and report the time scale of glitch recovery, if measured. Finally, in the last column we present the FMI due to the glitch using the Equation 11. of \citet{basuFMI}. We quote the 2$\sigma$ uncertainties in our measurement.}
\label{glitchparametertable}
\end{table*}
%We also report the largest glitch ($\Delta \nu/\nu = 3.1479 \times 10^{-6}\, \text{and} \, 8.374 \times 10^{-7}$) so far in PSR J1731$-$4744 and in PSR J0534+2200 respectively.
The observations of pulsars at the uGMRT  were done with a typical cadence of 20 days (Table \ref{Obs_tab}), whereas those at the ORT were performed with  much higher cadence (typically 2$-$3 days). The details of observations are listed in Table \ref{Obs_tab}. While our sample consisted of 18 pulsars, only the pulsars, where a glitch was detected are listed in the Tables \ref{Obs_tab} and \ref{glitchparametertable}.

Since data from both the uGMRT and the ORT were available in PSRFITS format, all subsequent analysis used the PSRCHIVE  package\footnote{http://psrchive.sourceforge.net/} \citep{hotan04}. First, a high S/N ratio profile was selected and modelled as a sum of Gaussians using the program \say{paas} to obtain a noise-free template profile. Then, this template was cross-correlated in the frequency domain with all profiles using \say{pat} employing a frequency domain cross correlation method developed by \cite{taylor1992}. The ToAs thus obtained were analysed using the pulsar timing software TEMPO2\footnote{https://bitbucket.org/psrsoft/tempo2/} \citep{hem06,ehm06}.Throughout this work, we used  the DE405 solar system ephemeris \citep{sta98} with the 2014.11.1 version of TEMPO2. The Earth rotation parameters for TEMPO2 were updated till 2019. The timing analysis for determining the glitch parameters used J2000 equatorial coordinates for the pulsars as shown in Table \ref{Obs_tab} along with the relevant references.

TEMPO2 provides the differences between actual pulse arrival times and times predicted from a simple assumed rotational model. These differences or residuals describes deviations from the assumed rotational behaviour of the pulsar. A spin-up, as observed in a glitch, is seen as an increasingly negative residual. A simple slow-down model involving rotational frequency and its first derivative is first fitted  to a period of time, which is devoid of any glitch activity. The timing residuals for the whole data set are then inspected visually for the presence of glitches. The fractional spin up $\Delta \nu/\nu$ and  fractional spin up rate $\Delta \dot{\nu}/\dot{\nu}$ were obtained from comparison of parameters before and after the glitch, with the  epoch of the glitch  determined by requiring a continuity of phase across the glitch. These parameters are listed for 11 glitches in Table \ref{glitchparametertable}.

To obtain the variation of $\nu$ as a function of epoch, first ToAs of all sub-integrations (typically 3 to 10) in each observation was obtained. Then, a subset of residuals were progressively chosen using a moving window of about 5 to 50 days and a stride of 3 to 30 days depending on the cadence of the observations and a simple slow-down model of the form $\nu(t) = \nu_0 + \dot{\nu}_0 \times (t-t_0)$ was fitted at the central epoch ($t_0$) of the moving window, where $\nu_0 \, \text{and} \, \Dot{\nu}_0$ are the measured rotation frequency and its derivative at the time $t_0$. This resulted in a time-series of $\nu$ as shown in the lower panels of Figure \ref{plotJ0534large} $-$ \ref{1837}. In few glitches the time-series of $\dot{\nu}$ can also be obtained as shown in Figure \ref{plotJ0534large}, \ref{plotJ0534small_late} and \ref{plotJ1740gmrt}.  The observed post-glitch frequency residuals can be described as a function of the time  elapsed since the epoch of the glitch, relative to the pre-glitch ephemeris, by different models, some of which are discussed in the next section.

\label{obsandanalysis}

\section{Results}
\label{results}
%\subsection{Pulsar Glitches detected from ORT}

\begin{figure}
    \centering
    \includegraphics[scale=0.4]{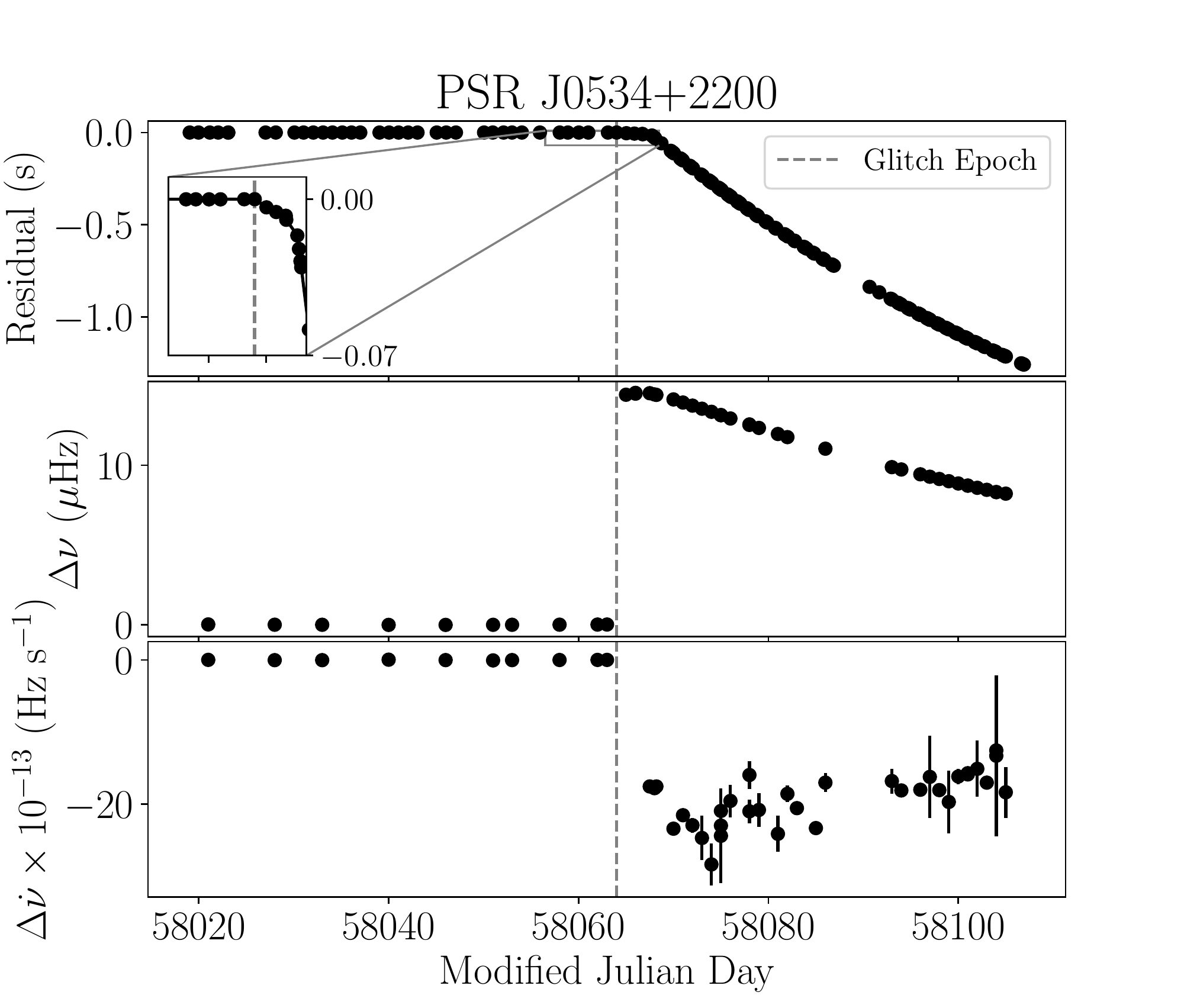}
    \caption{The largest ever glitch observed in PSR J0534+2200 (Crab pulsar) on 2017 November 7. While the spin up started at this epoch, the recovery was delayed by approximately 1.78 days. The top panel shows the evolution of timing residuals with observations epoch. The delayed recovery part in the timing residuals is shown in the box zoomed in near the glitch epoch in the upper panel. The middle and bottom plot show the evolution of $\nu$ and $\dot{\nu}$ respectively after subtracting the pre-glitch trends in these parameters to bring out the departures from a simple spin down model more clearly. The glitch epoch is marked with grey colour vertical line.}
    \label{plotJ0534large}
\end{figure}

\begin{figure}
    \centering
    \includegraphics[scale=0.4]{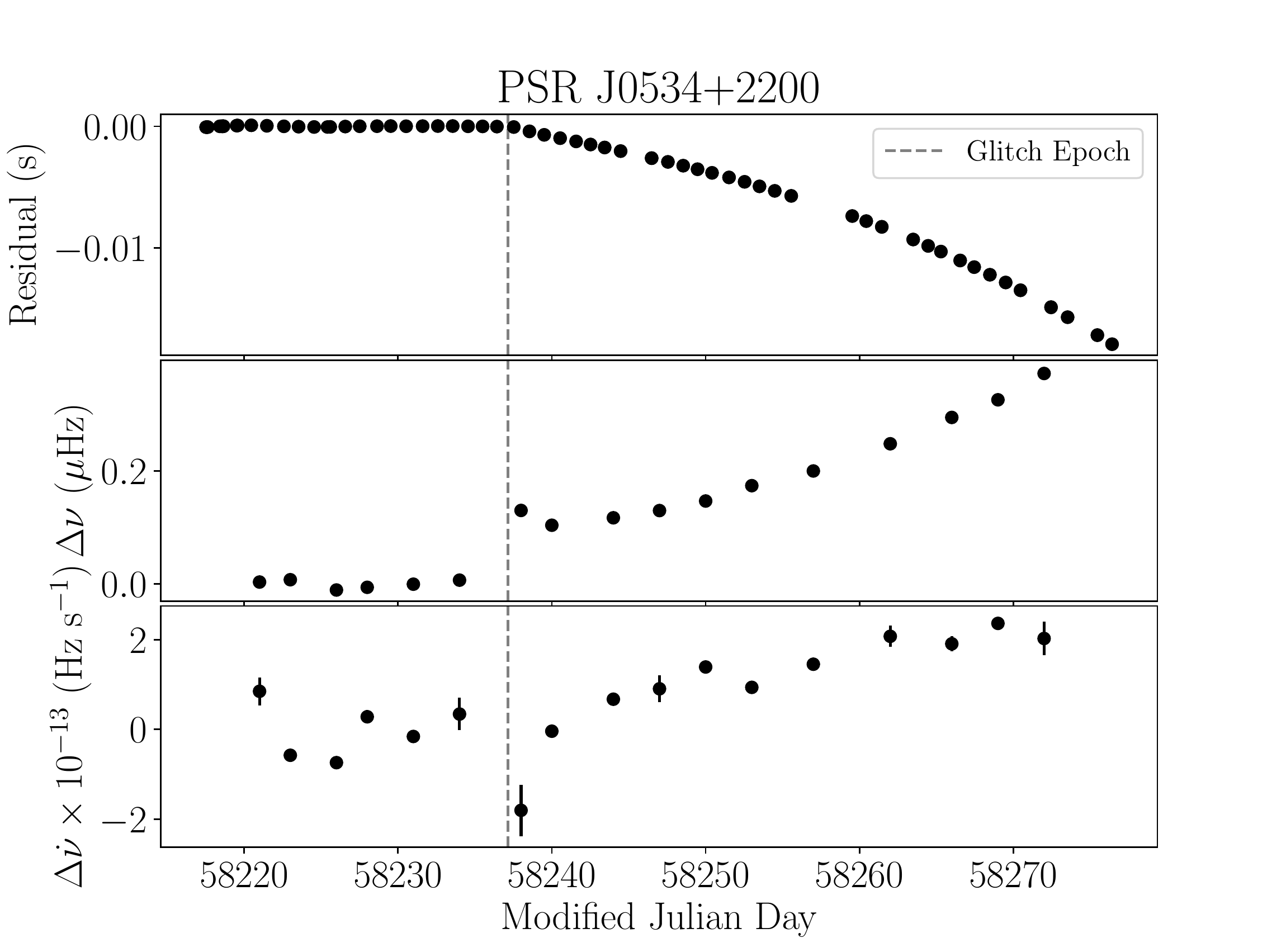}
    \caption{A smaller glitch at MJD 58237 succeeding the large glitch in PSR J0534+2200. The panels in the figure are similar to Figure \ref{plotJ0534large}}
    \label{plotJ0534small_late}
\end{figure}

\begin{figure}
    \centering
    \includegraphics[scale=0.4]{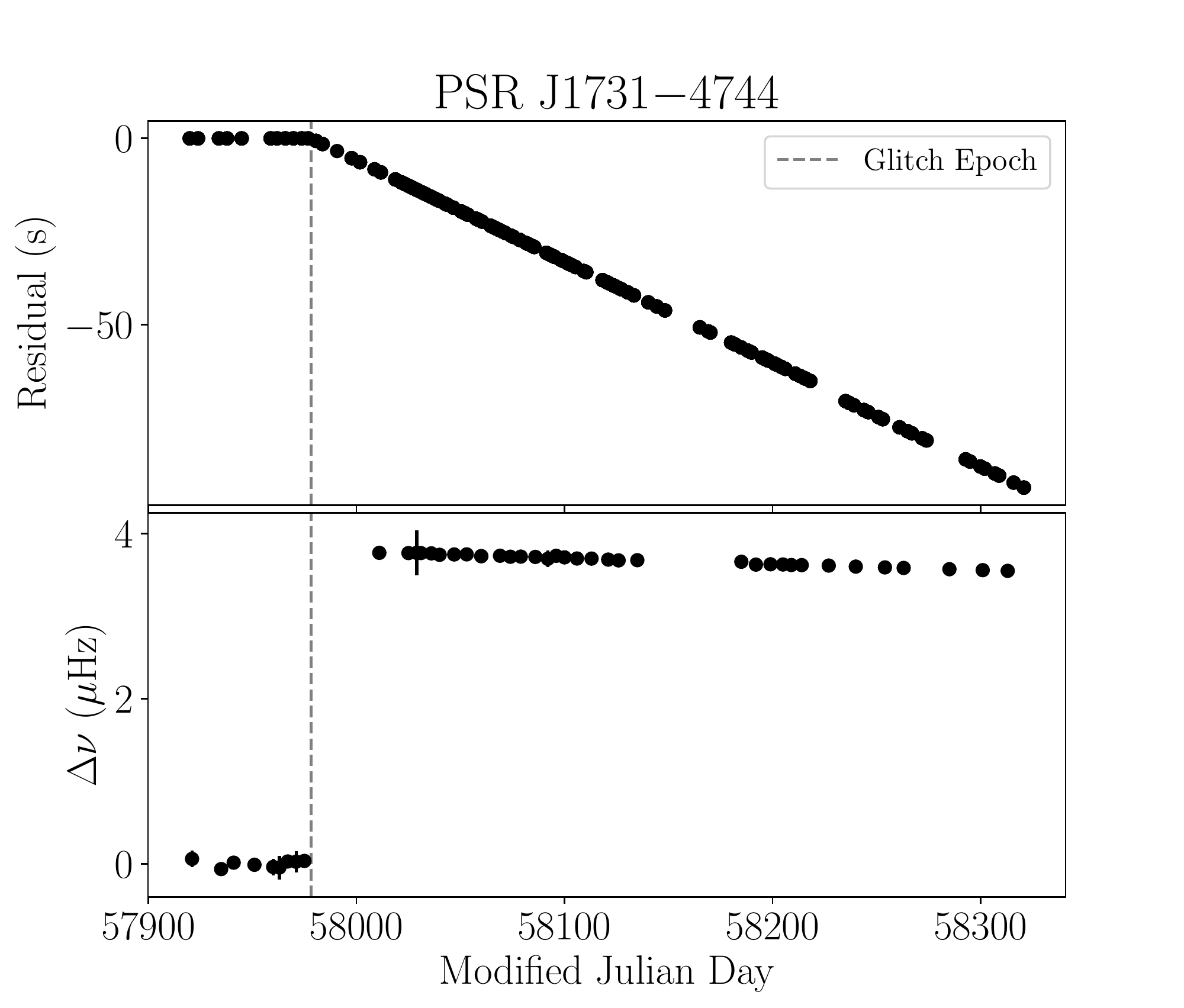}
    \caption{The time and frequency residuals for the glitch in PSR J1731$-$4744 on MJD 57978. The panels in the figure are similar to Figure \ref{plotJ0534large}. The frequency residuals were obtained by fitting $\nu$ only. Hence, $\Delta \dot{\nu}$ evolution is not shown.}
    \label{plotJ1731}
\end{figure}

In our monitoring, we detected 11 glitches from  8 pulsars. The parameters of these glitches are presented in Table \ref{glitchparametertable}. We present three new glitches in PSRs J0729$-$1448, J1751$-$3323 and J1837$-$0604. Two of the glitches, one each in PSR J0534+2200 and J0835$-$4510, reported in this paper were previously published \citep{shaw18,pdh+18}. We present new analysis of these glitches using  our data at 334.5 MHz with the ORT. For the remaining six glitches, which are also listed in the Jodrell Bank glitch catalogue, the glitch parameters obtained from our analysis as well as their post-glitch behaviour is presented for the first time. The post-glitch behaviour of all the pulsars is shown in Figure \ref{plotJ0534large} to  \ref{1837}, where the phase residuals, evolution of frequency and its derivative after subtracting the pre-glitch timing model as a function of observations epoch are shown. Evolution of $\dot{\nu}$ is shown only if $\dot{\nu}$ is fitted.A discussion on individual glitches is presented below.

\subsubsection{\textbf{PSR J0534$+$2200}}
\label{crab}
The largest ever glitch event in PSR J0534+2200 (Crab Pulsar) on MJD 58064 \citep{slb+17} was detected using the ORT \citep{kjbm17}. The glitch shows a feature, where the frequency offset from a secular spin down just after the glitch increases slowly for an initial two days, followed by the usual recovery as shown in Figure \ref{plotJ0534large}. We also detected an "aftershock" glitch about 175 days later, although this interval is much larger than 20 to 30 days as proposed by \cite{wbl01} for the aftershock glitches in their sample.

To understand post-glitch behaviour of glitches presented in this paper, we tried different phenomenological models. While this is particularly relevant for glitches in PSR J0534+2200, where a simple slow down model does not adequately describe the rich features seen in its glitches, different models may need to be adapted for other pulsars as well. Hence, we first define these models. A general phenomenological model for a glitch is given in Equation \ref{model1}.

\begin{eqnarray}
\label{model1}
    \Delta \nu (t \ge t_g) = \Delta \nu_p + \Delta \dot{\nu}_p(t-t_g) + \sum_{i=1}^{i=n} \Delta \nu_{d_i}e^{-\frac{t-t_g}{\tau_{d_i}}} ~~~~~\\
    \text{(MODEL$-$I)} ~~~~\nonumber
\end{eqnarray}
where $\Delta \nu_p$ is the permanent change in the rotational frequency, $\Delta \dot{\nu}_p$ is the permanent change in the frequency derivative and $t_g$ is the glitch epoch. In few cases \citep{espinoza2011}, the post-glitch epochs can be seen with multiple exponential recoveries, each with a time-scale $\tau_{d_i}$, with the amplitude of the decaying component given by $\Delta \nu_{d_i}$. However, for the Crab pulsar, it becomes necessary to account for the change in $\Ddot{\nu}$ \citep{wbl01}. Incorporating this term, therefore Equation \ref{model1} can be written as
\begin{eqnarray}
\label{model2}
    \Delta \nu (t \ge t_g)= \Delta \nu_p + \Delta \dot{\nu}_p(t-t_g) + \frac{1}{2}\Delta \Ddot{\nu}_p(t-t_g)^2 ~~~~~~~\\
   +\sum_{i=1}^{i=n} \Delta \nu_{d_i}e^{-\frac{t-t_g}{\tau_{d_i}}} ~~~~~~~~~~~~~~~~~~~~~~~~~~~~\nonumber  
   \\ \text{(MODEL$-$II)} \nonumber
\end{eqnarray}
It may be noted that $\Delta \Ddot{\nu}$ and the exponential terms become covariant for very large value of $\tau_d$. This is because the expansion of exponential function will have a constant term, a first order term and a quadratic term in time. For very large values of $\tau_d$, the expansion of the exponential series can be terminated at the second order term, absorbing the coefficients with $\Delta \nu_p, \Delta \dot{\nu}_p \, \text{and} \, \Delta \Ddot{\nu}_p$ terms.  This gives a third phenomenological model for the post-glitch behaviour given by Equation \ref{model3}.

\begin{eqnarray}
\label{model3}
    \Delta \nu (t \ge t_g)= \Delta \nu_p + \Delta \dot{\nu}_p(t-t_g) + \frac{1}{2}\Delta \Ddot{\nu}_p(t-t_g)^2
    \\ \text{(MODEL$-$III)}  \nonumber
\end{eqnarray}

Some pulsars show a post-glitch rotational behaviour which can be best described by two other subsets of the above phenomenological models given by Equations \ref{model4} and \ref{model5}.

\begin{equation}
    \label{model4}
    \Delta \nu (t \ge t_g)= \Delta \nu_p + \sum_{i=1}^{i=n} \Delta \nu_{d_i}e^{-\frac{t-t_g}{\tau_{d_i}}} ~~~~~~~~~~~~~\text{(MODEL$-$IV)}
\end{equation}

\begin{equation}
 \label{model5}
    \Delta \nu (t \ge t_g)=\sum_{i=1}^{i=n} \Delta \nu_{d_i}e^{-\frac{t-t_g}{\tau_{d_i}}} ~~~~~~~~~~~~~~~~~~\text{(MODEL$-$V)}
\end{equation}

\begin{figure}
    \centering
    \includegraphics[scale=0.4]{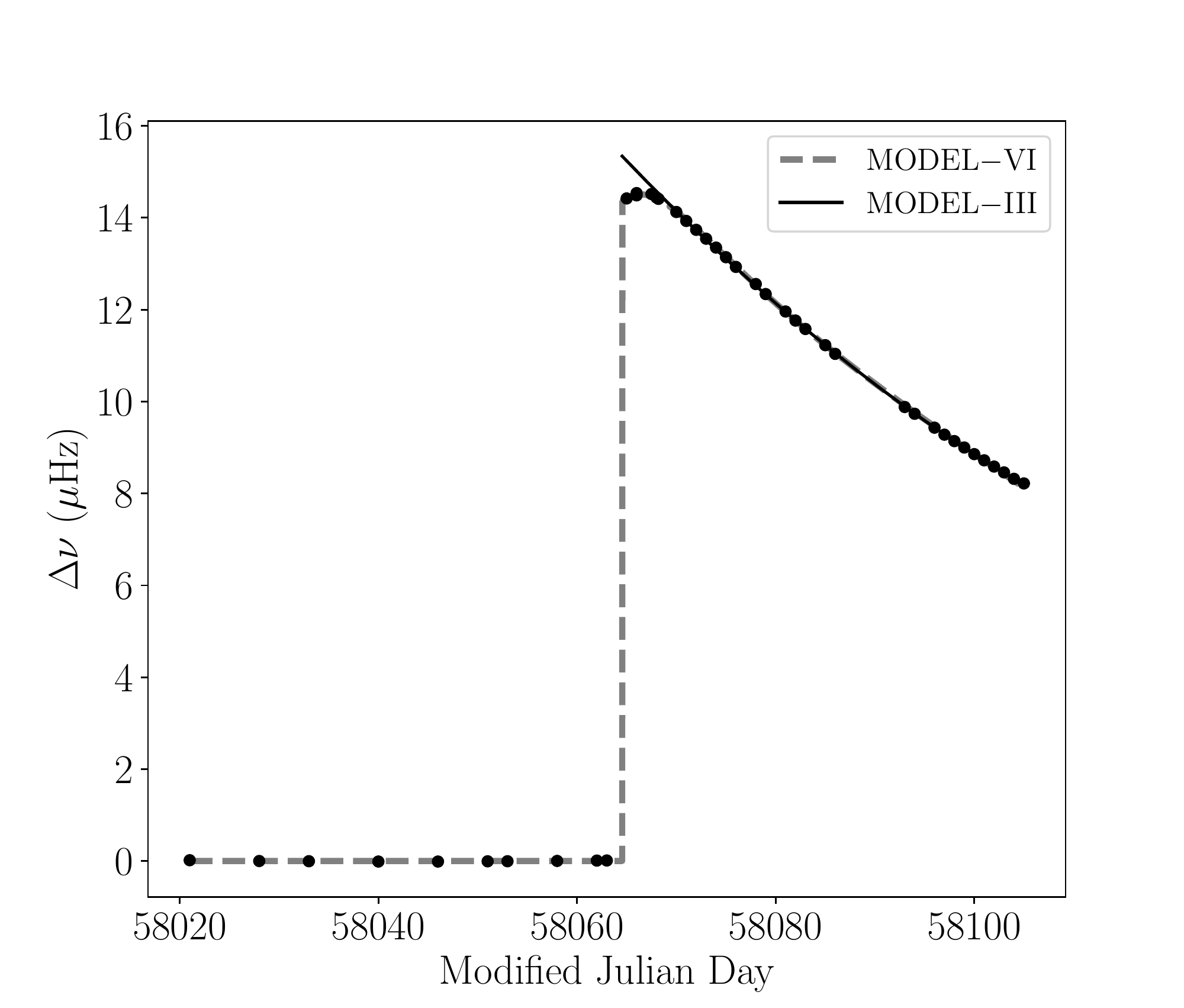}
    \caption{Modelled curve using MODEL$-$VI (Equation \ref{dampedevol}) is over-plotted on the post-glitch evolution of $\nu$ after subtracting the pre-glitch trends in this parameter for the large glitch in PSR J0534+2200 on MJD 58064. The best model was arrived using a Bayesian analysis.}
    \label{modelcrabplot}
\end{figure}

The post-glitch behaviour of the largest glitch in PSR J0534+2200 on 2017 November 7 (MJD 58064.9) shows a delayed recovery preceded by a slow increase in spin up over 1.78(1) days, which cannot be described by any of the above phenomenological models. Hence,  we propose a phenomenological model invoking a strongly damped oscillatory fast decaying transient response superposed on canonical post-glitch behaviour of this pulsar. This model is given by Equation \ref{dampedevol}, where $\Delta \nu_o$ is the amplitude of the oscillation with angular frequency $\Omega$, which get exponentially damped by a time scale denoted by $\zeta$.

\begin{eqnarray}
\label{dampedevol}
    \Delta \nu (t \ge t_g)= \Delta \nu_oe^{-\frac{t-t_g}{\zeta}} sin[\Omega(t-t_g)]+ \Delta \nu_p + \Delta \dot{\nu}_p(t-t_g) \\ 
     + \frac{1}{2}\Delta \Ddot{\nu}_p(t-t_g)^2 + \sum_{i=1}^{i=n} \Delta \nu_{d_i}e^{-\frac{t-t_g}{\tau_{d_i}}} \nonumber
     \\
     \text{(MODEL$-$VI)} \nonumber
\end{eqnarray}

From now onwards, we will refer to models described by Equations \ref{model1}, \ref{model2}, \ref{model3}, \ref{model4}, \ref{model5} and \ref{dampedevol}  as MODEL$-$I, MODEL$-$II, MODEL$-$III,  MODEL$-$IV, MODEL$-$V and MODEL$-$VI respectively. Wherever appropriate, different models can be tested against the data. In what follows, we compare fits to these models for PSR J0534+2200. We carried out best fits to MODEL-I to MODEL-V, with similar initial guess values for the common parameters for the large glitch at MJD 58064. It may be noted that $\Delta \nu$ shows  a slow rise and a turnaround in the initial two days after the glitch, which cannot be fitted by these models. Therefore, we  shift the origin $t_g$ to MJD 58070 to get better estimates of the parameters of the models with the assumption that the immediate transient response of the glitch decays down in the first 6 days after the glitch. From our fits, we find that compared to Model$-$I,II,IV and V, MODEL$-$III explain the post-glitch recovery the best.
However, this model only describes the long term delayed post-glitch behaviour and does not characterise the transient at the glitch epoch. To incorporate the transient along with the long  term behaviour, we used MODEL$-$VI. Since there are very few number of data points in the turn around region, we did a Bayesian analysis instead of least squares fit. We use EMCEE\footnote{https://emcee.readthedocs.io/en/v2.2.1/} \citep{EMCEE} package for running the Monte Carlo chains. The Bayesian analysis yields a fit shown in Figure \ref{modelcrabplot} with estimates for the parameters $\Delta \nu_o, \, \zeta, \, \Omega, \, \Delta \nu_{p},\,  \Delta \dot{\nu}_{p} \, \text{and} \, \Delta \Ddot{\nu}_{p}$ with reasonably constrained posterior distributions (not shown). The validity of the MODEL$-$VI over MODEL$-$III has been tested by computing the Bayesian Information Criterion (BIC). The BIC for a given model is given by Equation \ref{bic} \citep{Liddle2017}
\begin{equation}
\label{bic}
    \text{BIC} = -2\text{ln} \mathcal{L}_{max} + k\text{ln}N
\end{equation}
where $\mathcal{L}_{max}$ is the maximum likelihood, $k$ is the number of parameters and $N$ is the number of data points. A model with lower BIC is strongly preferred over the model with higher BIC if the difference between the BICs is greater than 10 \citep{kauss95}. We compute the $\Delta$BIC = BIC$_{\text{MODEL}-\text{III}}$ $-$ BIC$_{\text{MODEL}-\text{VI}}$ taking all the data points in the post-glitch epoch. We obtain $\Delta$BIC $=$ 801510, which implies the MODEL$-$VI is significantly favoured over MODEL$-$III. This is also visually evident from the fit shown in the Figure \ref{modelcrabplot}. However, the MODEL$-$VI is similar to 
%has a contribution from 
the MODEL$-$III at the later phase of the post-glitch evolution. Therefore, asymptotically the contribution towards post-glitch evolution in MODEL$-$VI should be similar to that in the MODEL$-$III. Hence, we perform the same test on the data points after MJD 58080. The measured $\Delta$BIC $=$ 153, suggesting that  the difference in the models is significantly smaller in this regime. Overall,  MODEL$-$VI is strongly preferred over MODEL$-$III. We quote the median of the posterior distribution as the best fit value for the parameters in Table \ref{fitmodel6} and quote $1\sigma$ error on the parameters obtained from the posterior distribution. The initial guess for the $\Delta \nu_{p}, \, \Delta \dot{\nu}_{p} \, \text{and} \, \Delta \Ddot{\nu}_{p}$ for MODEL$-$VI were taken from the best fit parameters of MODEL$-$III. While MODEL$-$VI is remarkably better than the other models, we still advise caution in using the MODEL$-$VI parameters, as the transient behaviour is constrained by only 2 degrees of freedom. A much higher cadence during a future large glitch in this pulsar is needed to validate this model significantly. 

\begin{table}
\begin{tabular}{ll}
\hline \hline
Parameter & Values              \\ \hline \hline
$\Delta \nu_p$         & 14.3562(7)$\times 10^{-6}$ Hz    \\ \hline 
$\Delta \dot{\nu}_p$        & -1907(2)$\times 10^{-15}$ Hz s$^{-1}$ \\ \hline
$\Delta \Ddot{\nu}_p$         & 70(1)$\times 10^{-21}$ Hz s$^{-2}$   \\ \hline
$\Delta \nu_o$         & 11.29(3)$\times 10^{-6}$ Hz\\ \hline
$\zeta$         & 5.16(1) days        \\ \hline
$P = \frac{2\pi}{\Omega}$        & 200(14) days        \\ \hline \hline
\end{tabular}
\caption{Table containing the values of the parameters of MODEL$-$VI fitted on the Crab pulsar glitch on MJD 58064.}\label{fitmodel6}
\end{table}

MODEL$-$VI can be used to arrive at an alternative estimate of the glitch epoch. We use the parameters from MODEL-VI to reconstruct 
the glitch event with 85 uniformly spaced data points as shown by the grey dashed line in Figure \ref{modelcrabplot}. We use this as a 
template, M(t), to estimate the glitch epoch. Similarly, the data  were interpolated to 85 points between MJD 58021 and 58105 giving a time 
resolution of 1 day, which we call as interpolated glitch data $G(t)$. The time resolution is equal to the gap between the last measured pre-glitch epoch and the first measured post-glitch epoch. The interpolated data were added with Gaussian random noise with root mean square as 6.6$\times 10^{-9}$ Hz, estimated from the uncertainties in the frequency measurement shown in the middle plot of Figure \ref{plotJ0534large}. 
%We place  M(t) with MJD 58064.555 at the zeroth time bin, where MJD 58064.555 was the glitch epoch quoted by \cite{shaw18}. to be . 
We then follow Taylor's Method \citep{taylor1992} to compute the time-of-arrival of glitch or the glitch epoch by minimizing the chi-square between $M(t)$ and $G(t)$ in the frequency domain. From the above mentioned method, we find the glitch epoch to be  MJD 58064.9(1), where this epoch now incorporates the slow rise seen in $\Delta \nu$ in Figure \ref{plotJ0534large}.  We quote the values of relative change in the rotation frequency and its first order derivative at this epoch in Table \ref{glitchparametertable}. The instantaneous fractional change in rotation frequency is 7 percent less with our method than that quoted by \cite{shaw18} as we use an epoch incorporating both the transient and delayed recovery post-glitch behaviour as well as consider the relatively smaller instantaneous change in spin parameter due to rising behaviour of $\Delta \nu$ at the glitch instant. In this model, the amplitude of oscillating component ($\Delta \nu_o$) is almost 79 percent that of the permanent change ($\Delta \nu_p$) in the rotation rate, although it is quickly damped by the large damping factor ($\zeta$) (small damping time scale of about 40 times smaller than oscillation period). We use the same model to compute the rise time of the glitch. From the middle panel of  Figure \ref{plotJ0534large}, it is evident that there exists a unique maximum. Therefore, we define $t_r$ as the rise time of the glitch, where the derivative of Equation \ref{dampedevol} vanishes. The derivative is independent of the permanent change in the rotational frequency $\Delta \nu_p$ as given in Equation \ref{derivative}.
\begin{eqnarray}
\label{derivative}
\Delta \nu_oe^{-\frac{t_r-t_g}{\zeta}}\{-\frac{1}{\zeta} sin[\Omega(t_r-t_g)] + \Omega cos[\Omega(t_r-t_g)]\}+ \\ \Delta \dot{\nu}_p
     +\Delta \Ddot{\nu}_p(t_r-t_g) = 0\nonumber
\end{eqnarray}
We solve the Equation \ref{derivative} by Brent's Method within 0 to 3 days from the glitch epoch. The measured value of rise time is 1.78(1) days. High cadence observations of a potential future large glitch in this pulsar would be important to verify if this is indeed the case. In addition to the large glitch, a small glitch was detected at MJD 58237. The spin up in this glitch is marginally higher than a similar \say{aftershock} glitch (glitch 10) reported by \citet{wbl01}. It is possible
that the star has not fully recovered from the previous large glitch. Hence, we used the post-glitch
model from the large glitch to estimate the parameters of this glitch. There is no apparent recovery
although the pulsar seems to exhibit a gradual spin up relative to pre-glitch rotational rate at late times post the glitch. However, this glitch is separated by about 175 days unlike the 20 to 30 day separation in the delayed \say{aftershock} reported by \citet{wbl01}, so it is unlikely to be an aftershock, particularly as no close succeeding glitch has been detected.

\subsubsection{\textbf{PSR J0729$-$1448}}
\begin{figure}
\centering
    \includegraphics[scale=0.4]{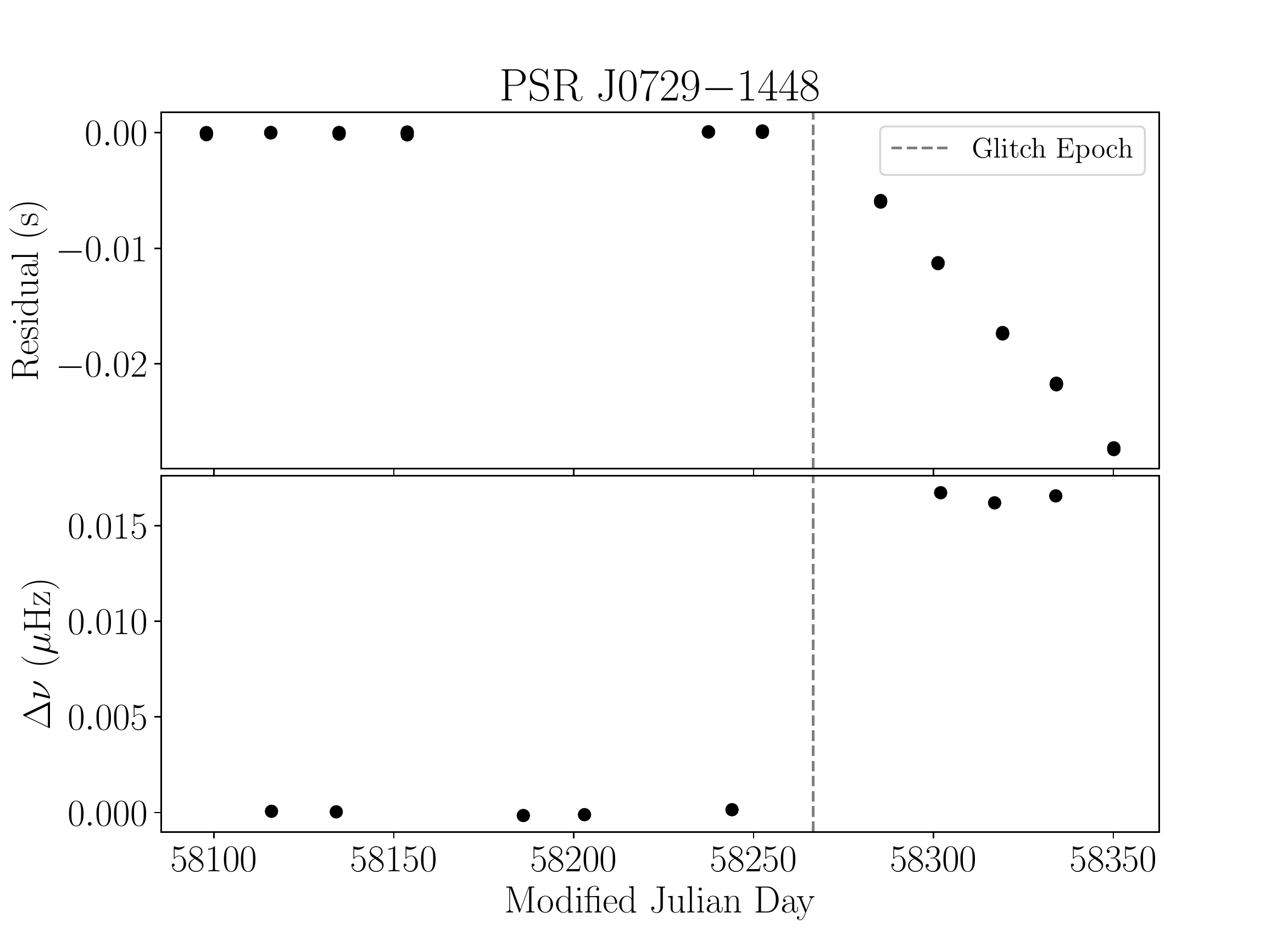}
    \caption{A small glitch in PSR J0729$-$1448 at MJD 58267. The panels in the figure are similar to Figure \ref{plotJ0534large}. Frequency residuals were obtained by fitting $\nu$ only. Hence, $\Delta \dot{\nu}$ evolution is not shown.}
    \label{plotJ0729}
\end{figure}

We report the sixth glitch in this pulsar. This new glitch is the smallest glitch observed so far and succeeds a three order of magnitude  larger glitch at MJD 54687. Unlike this large glitch, we do not see any appreciable change in frequency derivative. There seems to be a recovery just after the glitch, but our poor cadence near the glitch epoch does not allow us to distinguish this from timing noise in this pulsar, which may be of similar order. The time and frequency residuals are shown in Figure \ref{plotJ0729}.

\subsubsection{\textbf{PSR J0742$-$2822}}
\begin{figure}
    \centering
    \includegraphics[scale=0.4]{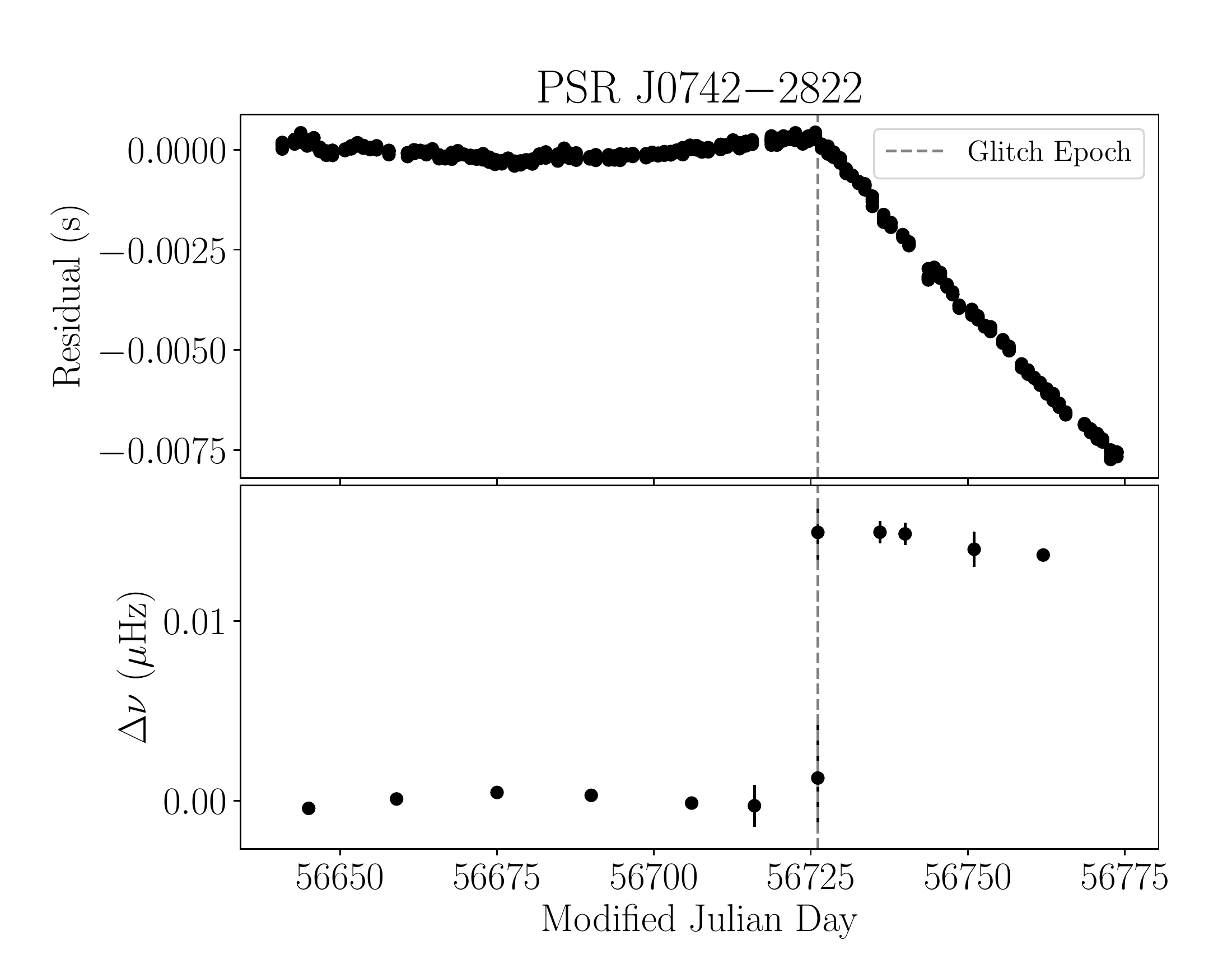}
    \caption{A small glitch in PSR J0742$-$2822 at MJD 56726. The panels in the figure are similar to Figure \ref{plotJ0534large}. Frequency residuals were obtained by fitting $\nu$ only. Hence, $\Delta \dot{\nu}$ evolution is not shown.}
    \label{plotJ0742}
\end{figure}

PSR J0742$-$2822 is a bright pulsar, which was monitored with a cadence of between 1 to 3 days at the ORT. The pulsar shows a large amount of timing noise, which may be related to its profile mode-change \citep{ksj13}. The frequency residuals were estimated using a moving box-car of width 15 days with a stride of 5 days. The higher cadence frequency measurement allow distinguishing the glitch from  the timing noise. A glitch was detected at MJD 56726 in the ORT data. The time and frequency residuals for this previously unpublished glitch are shown in Figure \ref{plotJ0742}.

\begin{figure}
    \centering
    \includegraphics[scale=0.4]{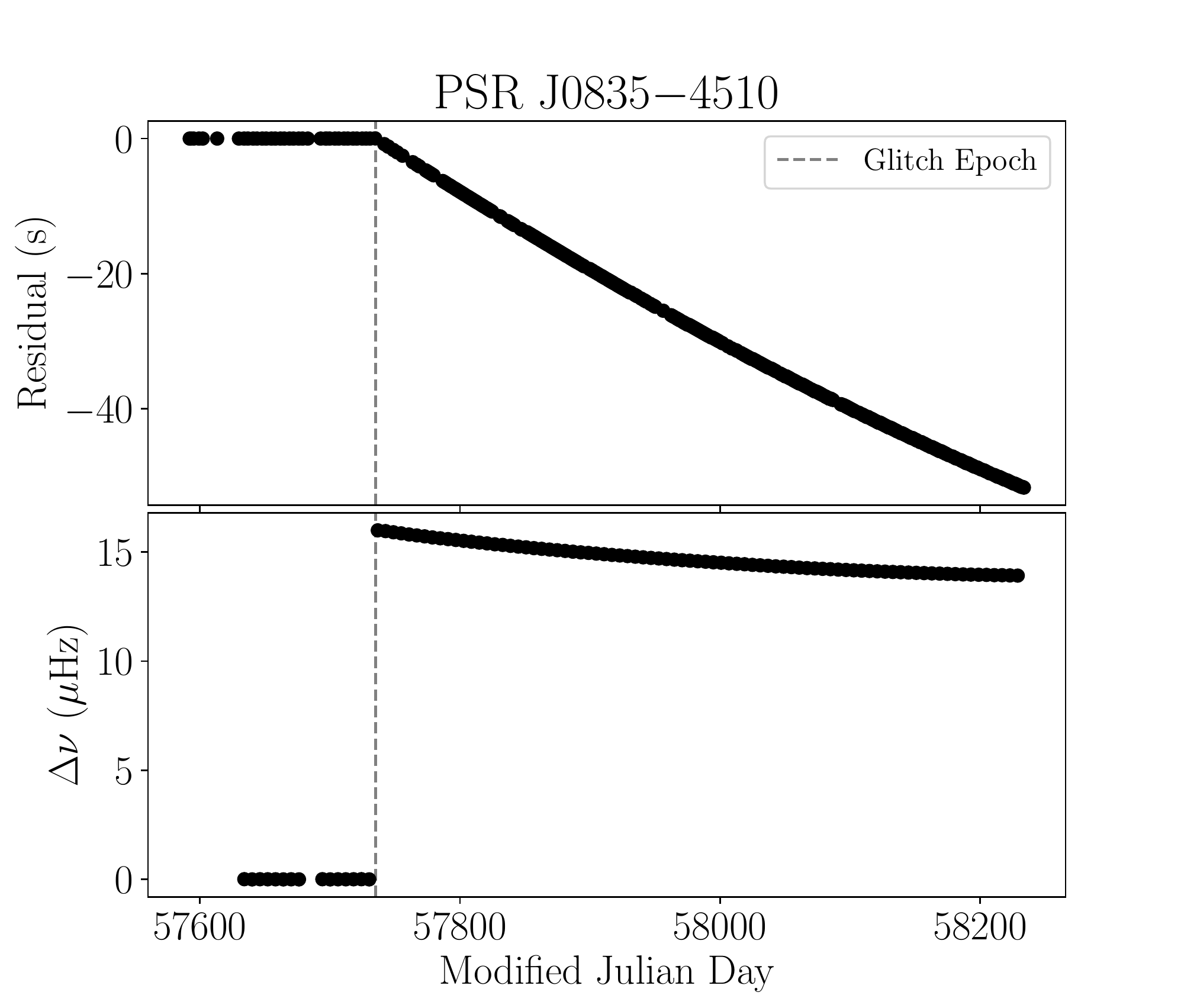}
    \caption{The time and frequency residuals for the glitch in PSR J0835$-$4510 on 2016 December 12. The panels in the figure are similar to Figure \ref{plotJ0534large}.Frequency residuals were obtained by fitting $\nu$ only. Hence, $\Delta \dot{\nu}$ is not shown.}
    \label{plotJ0835}
\end{figure}

\subsubsection{\textbf{PSR J0835$-$4510}}

The Vela pulsar (PSR J0835$-$4510) is monitored at the ORT with 1 to 3 day cadence. A large glitch in this pulsar was reported on 2016 December 12 \citep{pal16,pdh+18}. This glitch  was also detected at the ORT on MJD 57734. The glitch parameters are tabulated in the Table \ref{glitchparametertable} and the time and frequency residuals are shown in Figure \ref{plotJ0835}. The long term high cadence observations at the ORT allow following the rotational evolution of the pulsar with greater detail over 500 days. The post-glitch rotational evolution can be best described by MODEL$-$II. Like previous glitches in this pulsar, the pulsar rotation rate relaxes post-glitch with an exponential recovery. We did not have cadence smaller than a day unlike the monitoring reported elsewhere \citep{pdh+18,alg+19}, so we do not see fine scale features, such as pre-glitch spin-down, rise time and short 60 s exponential relaxation. But, we do see the long term exponential recovery with a  timescale $\tau_d$ of 32(2) days.  This timescale is comparable to those reported in previous glitches of this pulsar. The post-glitch evolution has a permanent changes in the rotational frequency $\Delta \nu_p = 15939.4(6)\times 10^{-9}$ Hz, along with the permanent change in the spin frequency derivative $\Delta \dot{\nu}_p =-8.00(3)\times10^{-14}$ Hz s$^{-1}$ and its second order derivative $\Delta \Ddot{\nu}_p = 1.52(1)\times10^{-21}$ Hz s$^{-2}$. Along with these permanent changes in the spin frequency and their higher order derivatives, we have measured a component of rotation frequency $\Delta \nu_d = 9.3(3)\times 10^{-8}$ Hz, which recovers exponentially.  

The healing parameter $Q$ of a glitch is defined as a fraction of the spin frequency recovered exponentially to the total change in the spin frequency. Therefore, the healing parameter can be written as $Q=\Delta \nu_d/(\Delta \nu_d + \Delta \nu_p)$ from MODEL$-$II. The measured healing parameter for this Vela glitch is $5.8\times 10^{-3}$.

\subsubsection{\textbf{PSR J1731$-$4744}}
A large proper motion has been reported for this pulsar in the literature. For our timing solution, we have used the proper motion computed by \citet{Skz+2019} assuming the proper motion in RA and DEC to be $\mu_{\alpha} = $ 63 mas yr$^{-1}$ and $\mu_{\delta}$ = -83 mas yr$^{-1}$ respectively. The J2000 position of the pulsar has been taken from the measurement made by \citet{Yu2013} as given in the Table \ref{Obs_tab}. This pulsar exhibited the largest glitch seen so far with a fractional spin up about 85 times that seen in previous large amplitude glitch on MJD 49387 \citep{espinoza2011}. The high cadence observations with the ORT allowed us to narrow down the glitch epoch to about a fraction of an hour by extrapolating the pre- and post-glitch phase. Our continued monitoring of the post-glitch phase of this pulsar shows no sign of an exponential recovery. However, the glitch is accompanied by a significant decrease in frequency derivative. The change in the frequency derivative was $\Delta \Dot{\nu}$ = 2.43 $\pm 0.06 \times 10^{-13}$ Hz s$^{-1}$.

\begin{figure}
    \centering
    \includegraphics[scale=0.3]{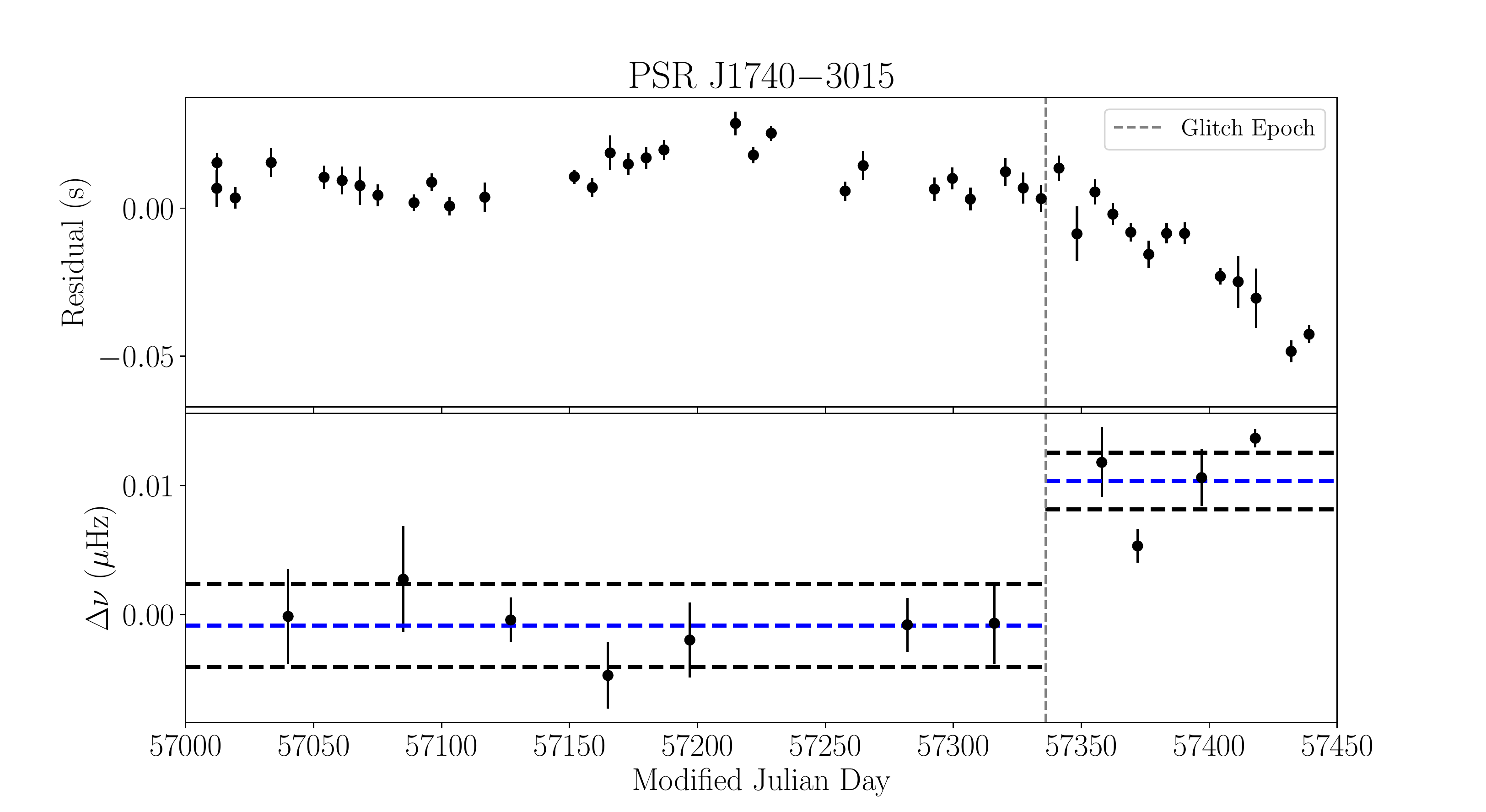}
    \caption{The time and frequency residuals for glitch 34 in PSR J1740$-$3015 on MJD 57336. The frequency residuals were obtained by fitting $\nu$ only. Hence, $\Delta \dot{\nu}$ evolution is not shown.}
    \label{plotJ1740_ort_small}
\end{figure}

\begin{figure}
    \centering
    \includegraphics[scale=0.4]{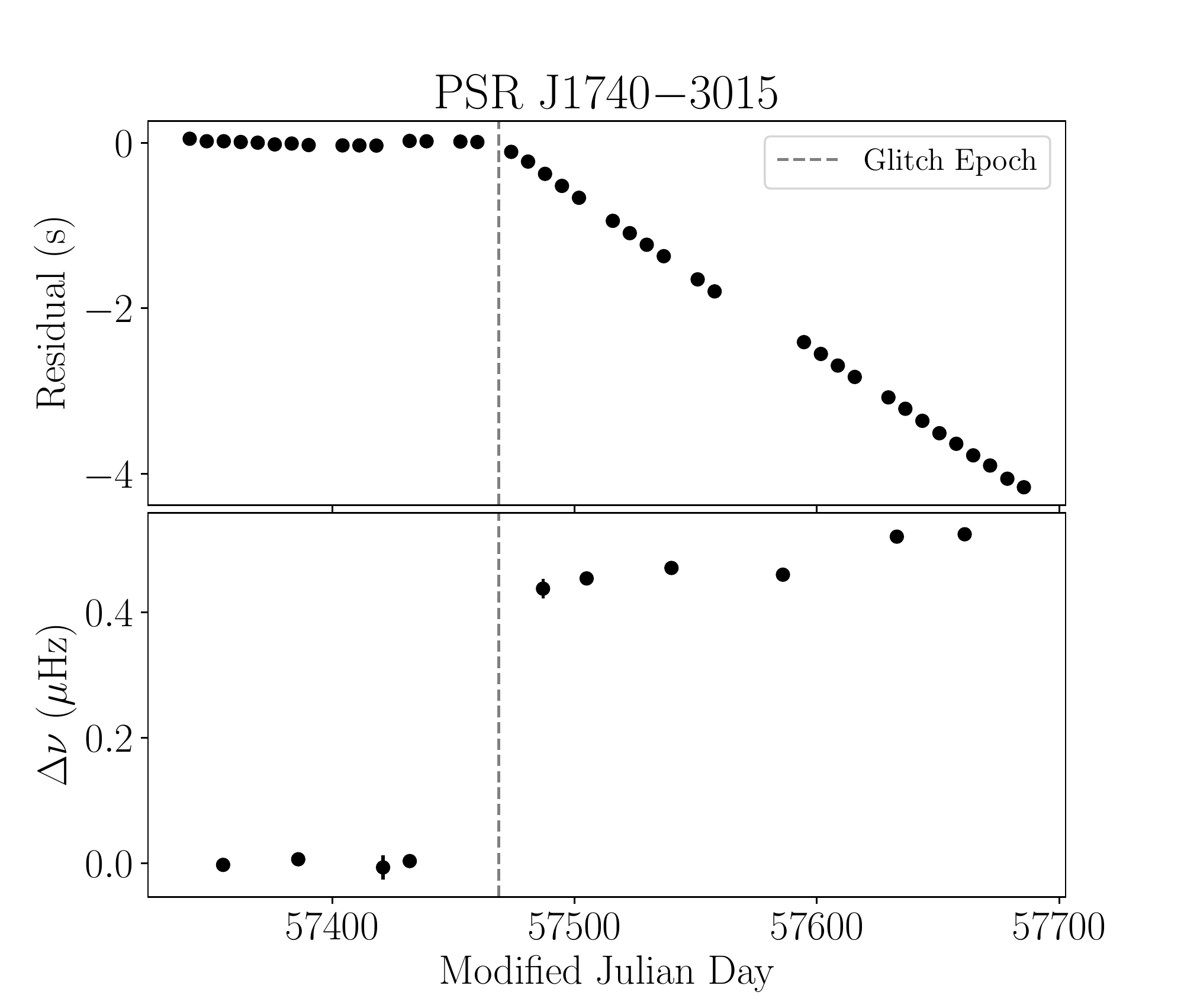}
    \caption{The time and frequency residuals for glitch 35 in PSR J1740$-$3015 on MJD 57468. The frequency residuals were obtained by fitting $\nu$ only. Hence, $\Delta \dot{\nu}$ evolution is not shown.}
    \label{plotJ1740}
\end{figure}

\begin{figure}
    \centering
    \includegraphics[scale=0.3]{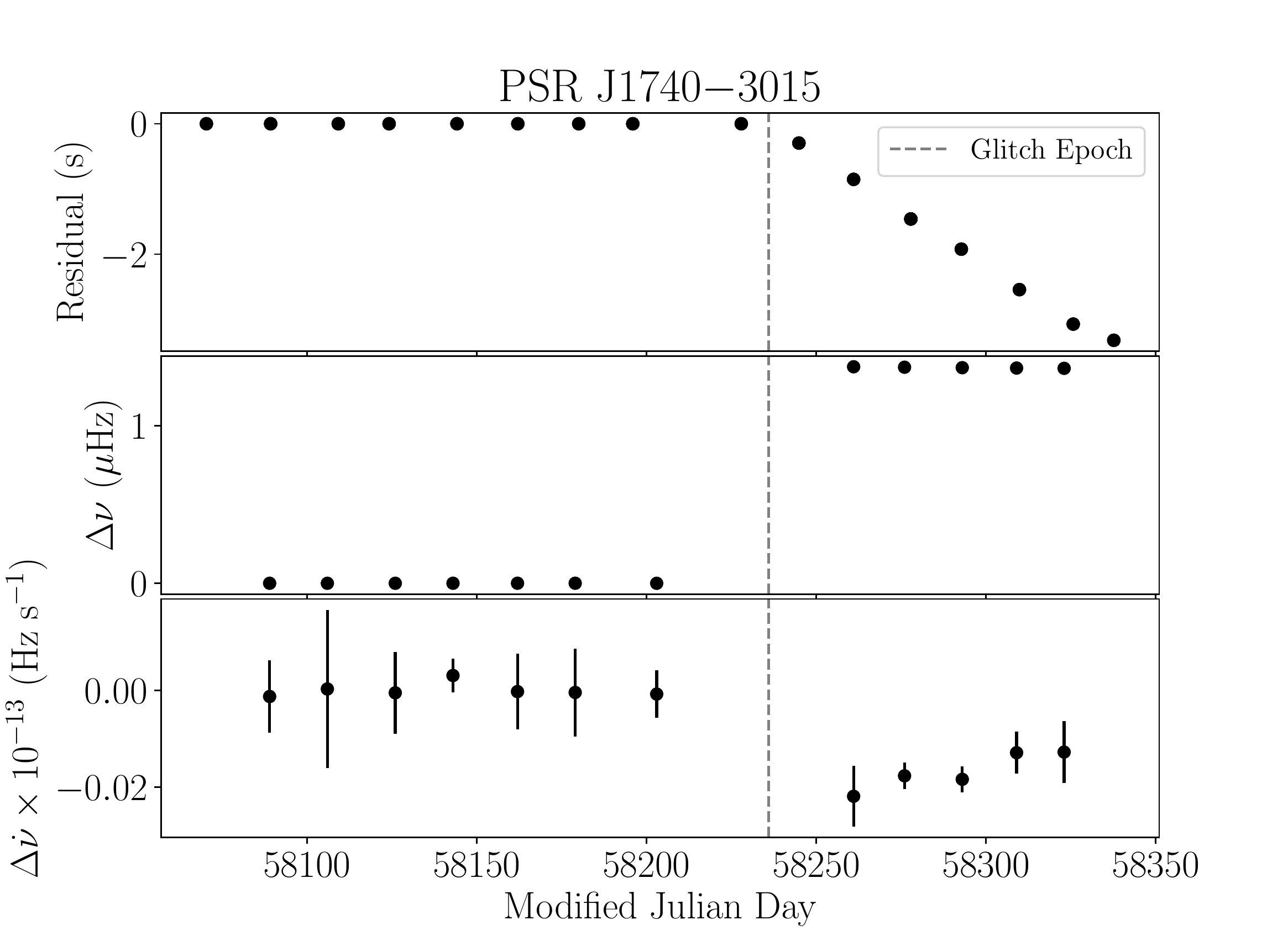}
    \caption{The time and frequency residuals for glitch 36 in PSR J1740$-$3015 on MJD 58236.}
    \label{plotJ1740gmrt}
\end{figure}

\subsubsection{\textbf{PSR J1740$-$3015}}
PSR J1740$-$3015 is one of the five pulsars with the largest number of glitches reported. It very often shows both large and small magnitude glitches. In our data spanning about 5 years, the pulsar glitched three times. The time and frequency residuals of these three glitches at MJD 57336, 57468 and 58236 are shown in Figures \ref{plotJ1740_ort_small}, \ref{plotJ1740} and \ref{plotJ1740gmrt} respectively. The former two were detected with the ORT and the last one was detected with the uGMRT. These glitches break the previous pattern shown by the pulsar, where a large dominant glitch was typically followed by several small glitches and the glitch amplitude of large glitches was progressively increasing as the fractional spin up in glitch 35 and 36 was about 235 $\times$ 10$^{-9}$ and 837  $\times$ 10$^{-9}$, much smaller than that in the previously reported large glitch on MJD 55213 \citep[Glitch 32 : fractional spin up 2668   $\times$ 10$^{-9}$][]{ymh+13}. No noticeable recovery is seen in the post-glitch behaviour. This is also made difficult by the small inter-glitch duration.

\subsubsection{\textbf{PSR J1751$-$3323}}

\begin{figure}
    \centering
    \includegraphics[scale=0.4]{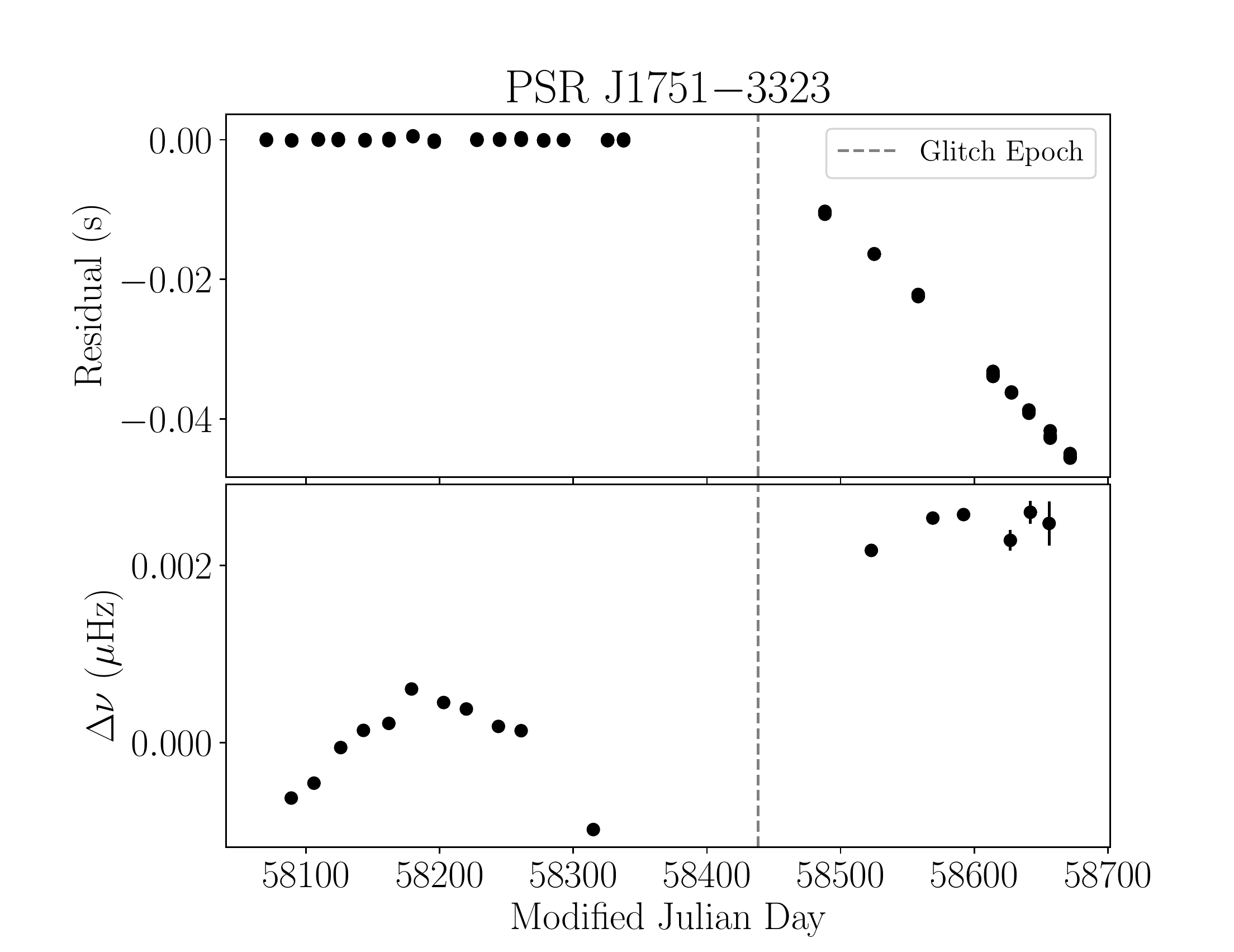}
    \caption{The time and frequency residuals for the new glitch  in PSR J1751$-$3323 on MJD 58438. The frequency residuals were obtained by fitting $\nu$ only. Hence, $\Delta \dot{\nu}$ evolution is not shown.}
    \label{plotJ1751}
\end{figure}

A new glitch was detected in PSR J1751$-$3323 with uGMRT on MJD 58438. The pulsar exhibits large amounts of timing noise in the inter-glitch interval as is evident from the frequency residuals shown in Figure \ref{plotJ1751}. This makes recognizing any recovery difficult. The amplitude of fractional rotational frequency change in this glitch is similar to that in the previous three reported glitches.

\subsubsection{\textbf{PSR J1837$-$0604}}

\begin{figure}
    \centering
    \includegraphics[scale=0.4]{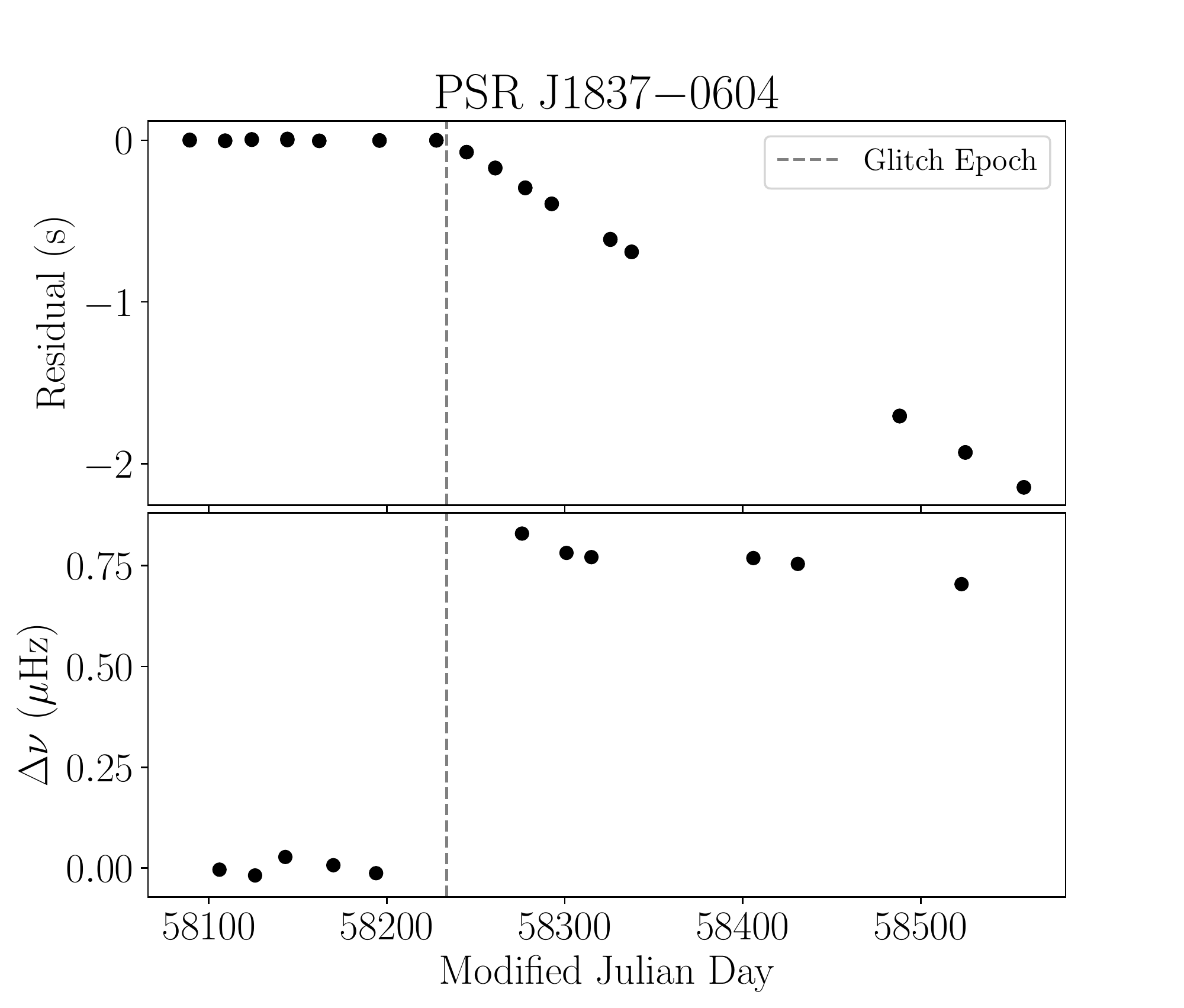}
    \caption{The time and frequency residuals for the new glitch  in PSR J1837$-$0604 on MJD 58233. The frequency residuals were obtained by fitting $\nu$ only. Hence, $\Delta \dot{\nu}$ evolution is not shown.}
    \label{1837}
\end{figure}

We detected a new glitch with uGMRT, the third and the second largest in this pulsar. There is no apparent short term recovery. Our data span was not long enough to characterize any change in the rotational derivative for this pulsar. The time and frequency residuals for this pulsar are shown in Figure \ref{1837}. There is a hint of post-glitch exponential recovery of about 20 days, but our cadence was too sparse to estimate this with a fit.

%\subsubsection{\textbf{PSR J1952$+$3252}}

\section{Discussions \& Conclusions}
\label{end}

Eleven glitches in 8 pulsars are presented in this paper. Three of these are being reported for the first time. The results include the largest ever glitch seen in the Crab pulsar on 2017 November 7 with a signature of a transient increase in the frequency residual for two days followed by the usual recovery seen in other large glitches in this pulsar. We present a re-analysis of this glitch with our high cadence data obtained using the ORT and show that the post-glitch behaviour require a model with a short term transient response and a longer term recovery involving the rotational frequency and its two higher order derivatives. The short term transient response may be described by an overdamped second order transfer function as one of the probable models. 
Assuming this model, the glitch epoch and the fractional spin-up is estimated to be 58064.9(1) and  484.39(1) $\times$ 10$^{-9}$ respectively. Reanalysis of glitch in the Vela pulsar on December 2016 using data from the ORT monitoring is also presented. High cadence long term monitoring with the ORT reveals an exponential relaxation time scale of 32(2) days with a healing factor Q = 5.8 $\times$ 10$^{-3}$. The sample of glitches presented in this work exhibit a fractional spin up ranging from 2$\times$10$^{-9}$ in Crab pulsar (PSR J0534+2200) to 3000 $\times$ 10$^{-9}$ (PSR J1731$-$4744). We also report the largest glitch $\Delta \nu/\nu = 3147.9 \times 10^{-9}$ so far in PSR J1731$-$4744. Post-glitch recovery is seen in PSR J0534+2200 and J0835$-$4510, whereas a hint of exponential recovery is visible in PSR J1837$-$0604. There is a significant decrease in frequency derivative in PSR J1731$-$4744.\\

The short term transient feature in the first two days following the largest glitch in PSR J0534+2200 seen in our ORT monitoring is consistent with similar  features reported in previous large glitches \citep{lsp92,wbl01}. This feature was also previosuly reported in observations of this glitch by Jodrell Bank Observatory \citep{shaw18}. A similar transient, albeit over a much shorter time scale 12.6 s, is also reported in high cadence observations of recent large glitch in the Vela pulsar \citep{pdh+18,alg+19}. These transient responses  point to a disturbance in the stable configuration of vortices and can allow for a probe of the dynamics of the participating components of the star as well as determine the coupling and mutual friction between the crustal and core superfluid components \citep{gca18,hka+18}. An important constraint for these models will be to explain both the short time transient as well as long term recovery within the same framework. This motivates continued high cadence monitoring of these two pulsars with the available telescopes. An automated way of triggering high cadence observations as soon as a glitch is detected is also required.
\\

Multi-element future telescopes, such as Square Kilometer Array (SKA), can be very useful for such high cadence monitoring, particularly if multiple subarrays are available. Such subarray mode of observations is planned for SKA. A smaller subarray of SKA Phase I can be used to carry out alternate day monitoring program of a sample of dozen Crab and Vela like pulsars, apart from observations of the Vela pulsar itself. Coupled with an automated glitch detection pipeline this can allow a larger number of such events to be characterized with pulsars with different spin histories. This can be potentially useful to understand the glitch dynamics as well as internal structure of neutron stars.

\section*{Acknowledgements}
We thank the staff of the Ooty Radio Telescope and the Giant Metrewave Radio Telescope for taking observations over such a large number of epochs. Both these telescopes are operated by National Centre for Radio Astrophysics of Tata Institute of Fundamental Research. PONDER backend, used in this work, was built with TIFR XII plan grants 12P0714 and 12P0716. We thank Dr. Jim Palfreyman for a careful and critical reading of the manuscript. BCJ, MAK acknowledge SERB grant.

%%%%%%%%%%%%%%%%%%%%%%%%%%%%%%%%%%%%%%%%%%%%%%%%%%



\bibliographystyle{mnras}
\bibliography{glitch_data_paper.bib}

%%%%%%%%%%%%%%%%%%%%%%%%%%%%%%%%%%%%%%%%%%%%%%%%%%
\iffalse
%%%%%%%%%%%%%%%%% APPENDICES %%%%%%%%%%%%%%%%%%%%%

\appendix

\section{Some extra material}

If you want to present additional material which would interrupt the flow of the main paper,
it can be placed in an Appendix which appears after the list of references.

%%%%%%%%%%%%%%%%%%%%%%%%%%%%%%%%%%%%%%%%%%%%%%%%%%

\fi
% Don't change these lines
\bsp	% typesetting comment
\label{lastpage}
\end{document}